\documentclass[twocolumn,prd,showpacs,
floatfix,preprintnumbers,nofootinbib]{revtex4}

\usepackage{epsfig}
\usepackage{amsmath,bm}

\newcommand{\I}{{\rm i}}

\graphicspath{{./figs/}}

\begin{document}

\title{Spurious instabilities in multiangle simulations of
collective flavor conversion}

\author{Srdjan Sarikas, David de Sousa Seixas and Georg Raffelt}
\affiliation{Max-Planck-Institut f\"ur Physik
(Werner-Heisenberg-Institut), F\"ohringer Ring 6, 80805 M\"unchen,
Germany}

\date{November 16, 2012}


\begin{abstract}
The dense neutrino flux streaming from a core-collapse supernova can
undergo self-induced flavor conversion caused by neutrino-neutrino
refraction. Numerical studies of these nonlinear effects are
challenging because representing the neutrino radiation field by
discrete energy and angle bins can easily lead to unphysical
solutions. In particular, if the number of angle bins $N_a$ is too
small, flavor conversion begins too deep and produces completely
spurious results. At the same time, $N_a=1$ (single-angle
approximation) can be a good proxy for the $N_a\to\infty$ limit.
Based on a linearized stability analysis, we explain some of the
puzzling effects of discrete angle distributions.
\end{abstract}

\preprint{MPP-2012-139}

\pacs{14.60.Pq, 97.60.Bw}

\maketitle

\section{Introduction}
\label{sec:intro}

The almost freely streaming flux of neutrinos emitted from a
collapsed supernova (SN) core starts at a radius of 10--30~km,
depending on the explosion phase and the neutrino energies, and at a
density of $10^{12\hbox{--}13}~{\rm g}~{\rm cm}^{-3}$. In
such conditions, neutrino refraction in matter is so large that the
propagation eigenstates are almost identical to the weak interaction
eigenstates, even though all neutrino mixing angles are relatively
large. Despite this simple boundary condition, the subsequent
neutrino flavor evolution is surprisingly complicated due to the
nonlinear impact of neutrino-neutrino refraction~\cite{Duan:2006an,
Duan:2010bg}.

One major problem in understanding this phenomenon is that numerical
studies are challenging. If the integration begins close to the
neutrino-emitting region of the SN (the ``neutrino sphere''), large
oscillation frequencies caused by the matter effect require many
time steps. Moreover, the neutrino radiation field must be
represented in some numerical form, in practice a finite number of
energy and angle bins. Thus far axial symmetry of the neutrino
stream was always assumed, implying that there was only one angle
variable. Even with this simplification, reliable results require a
large-scale computing effort~\cite{Duan:2008eb}.

Far away from the SN core, for example for a detector on Earth, the
angle distribution is irrelevant and good energy resolution is
sufficient. In particular, one can study the sharp spectral features
that self-induced flavor conversion can produce~\cite{Duan:2006an,
Raffelt:2007cb, Duan:2007fw, Fogli:2007bk, Dasgupta:2009mg}.
Therefore, many studies used the single-angle approximation, i.e.,
all neutrinos are put in a single angle bin corresponding to
emission at the neutrino sphere at one specific angle relative to
the radial direction. The simplest assumption of radially emitted
neutrinos is not possible because parallel-moving neutrinos do not
provide a refractive effect on each other. We always represent the
single-angle case by assuming neutrino emission at $45^\circ$ at the
neutrino sphere.

To go beyond single-angle studies, upgrading to a small number $N_a$
of angular bins is not enough. Beginning with the earliest
studies~\cite{Duan:2006an}, all multiangle simulations were haunted
by the same peculiar effect. Although simulations with a large number
of angular bins ($N_a\to\infty$) and those with just a single angle ($N_a=1$)
often provide similar results, simulations using a relatively small
$N_a$ lead to results completely different from the two limiting
cases. The required $N_a$ to avoid spurious solutions had completely
eluded explanation. Depending on circumstances, a few tens of modes
might be enough, whereas in other cases one needs thousands or more
bins.

For a simple (``single-crossed'' \cite{Dasgupta:2009mg}) neutrino
spectrum, which is relevant during the SN accretion phase,
self-induced flavor conversion begins at a critical onset radius. If
only neutrino-neutrino refraction is relevant, then at larger
neutrino densities, close to the SN core, the system is stable
(``sleeping top'' phase \cite{Hannestad:2006nj, Duan:2007mv}). The
onset radius turns out to be almost the same in the single-angle and
many-angle limits. On the other hand, if $N_a>1$ is too small,
flavor conversion begins at a smaller radius (where higher neutrino
densities dominate) and tends to cause kinematical decoherence among
angle modes~\cite{EstebanPretel:2007ec}. Similar spurious solutions
appear also when the normal matter effect is important.

This situation is to be contrasted with the role of energy bins that
can be chosen to the desired resolution without affecting the
qualitative behavior. Notice that neutrino energies (or rather the
vacuum oscillation frequencies) appear in the linear part of the
Hamiltonian, responsible for the vacuum oscillations. The angle
variables, on the other hand, appear in the neutrino-neutrino
interaction part which is the source of nonlinear effects. In the
single-angle multienergy case, the flavor evolution is described by
a small number of collective variables (``$N$-mode coherence''),
independently of the number of energy bins \cite{Raffelt:2010za,
Raffelt:2011yb, Yuzbashyan:2008}. The ultimate source for this
simple behavior is the simplicity of the single-angle Hamiltonian
which contains as many constants of the motion as variables and thus
is integrable~\cite{Raffelt:2011yb, Pehlivan:2011hp}. If detailed
energy resolution is not important, we can therefore study
conceptual aspects of multiangle effects in the monochromatic
approximation where we use only one energy bin for neutrinos and one
for antineutrinos.

The observed flavor conversion effect in the case of few angle modes
must begin with an instability, i.e., with a runaway flavor mode in
the interacting neutrino system. Understanding self-induced flavor
conversion based on a linearized stability analysis was pioneered by
Sawyer~\cite{Sawyer:2008zs} and further developed by our group and
collaborators~\cite{Banerjee:2011fj, Sarikas:2011am}. We here apply
this technique to the \hbox{``$N_a$ effect''} and find that many
puzzling numerical observations easily fall into place. The spectrum
of runaway modes is indeed very different for ``$N_a={}$few'' from
$N_a=1$ and $N_a\to\infty$. Additionally, the two neutrino mass
hierarchies show striking differences.

\section{Linearized stability approach}
\label{sec:linearized}

\subsection{Equations of motion}

Following earlier works~\cite{EstebanPretel:2008ni, Banerjee:2011fj,
Sarikas:2011am}, we describe the two-flavor neutrino field by
energy- and angle-dependent $2{\times}2$ matrices
${\bf\Phi}_{E,u}(r)$. Boldface characters denote matrices in flavor
space. The diagonal ${\bf\Phi}_{E,u}$ elements are the ordinary
number fluxes $F_{E,u}^\alpha$ (flavor $\alpha$) integrated over a
sphere of radius $r$, with negative $E$ for antineutrinos.
Moreover, we use the ``flavor isospin convention'' where
$F_{E,u}^\alpha<0$ for antineutrinos ($E<0$) and $F_{E,u}^\alpha>0$
for neutrinos ($E>0$).

The off-diagonal elements of ${\bf\Phi}_{E,u}(r)$, which are
initially very small, represent phase information caused by flavor
oscillations. The flavor evolution follows from the
Schr\"odinger-like equation \cite{EstebanPretel:2008ni}
\begin{equation}
{\rm i}\partial_r{\bf\Phi}_{E,u}=[{\bf H}_{E,u},{\bf\Phi}_{E,u}]
\end{equation}
with the Hamiltonian matrix
\begin{eqnarray}\label{eq:hamiltonian}
{\bf H}_{E,u}&=&\frac{1}{v_{u}}\,\left(\frac{{\bf M}^2}{2E}+
\sqrt{2}\,G_{\rm F}{\bf N}_\ell\right)\\
&+&\frac{\sqrt{2}\,G_{\rm F}}{4\pi r^2}\int_{-\infty}^{+\infty}dE'
\int_0^1 du'
\frac{1-v_{u}v_{u'}}{v_{u}v_{u'}}\,{\bf\Phi}_{E',u'}\,.
\nonumber
\end{eqnarray}
The matrix ${\bf M}^2$ of neutrino mass squares causes vacuum flavor
oscillations and that of net charged lepton densities ${\bf
N_\ell}={\rm diag}(n_e{-}n_{\bar
e},n_\mu{-}n_{\bar\mu},n_\tau{-}n_{\bar\tau})$ adds the matter
effect. The third term provides neutrino-neutrino refraction. A
neutrino radial velocity at radius $r$ is
$v_u=(1-u\,R^2/r^2)^{1/2}$, where $R$ is the radius of the neutrino
sphere, at which we label the neutrino angle modes by their emission
angle $\theta_R$. Our angle variable $u$ is then defined by
$v_u|_{r=R}=\cos\theta_R=(1-u)^{1/2}$, equivalent to
$u=\sin^2\theta_R$. The factor \hbox{$1-v_u v_{u'}$} comes from the
current-current nature of the weak interaction and leads to
multiangle effects. Moreover, $v_u$ appears in the denominator
because we follow the flavor evolution projected on the radial
direction.

Next, we study the instability driven by the atmospheric $\Delta
m^2$ and the mixing angle $\theta_{13}$. In the relevant SN region,
propagation eigenstates are almost identical with weak-interaction
eigenstates unless self-induced flavor conversion occurs. This means
that initially the off-diagonal elements of ${\bf\Phi}_{E,u}$ are
very small, justifying a linearized stability analysis, but
otherwise we do not need a specific numerical value of the mixing
angle.

We switch to the variable $\omega=\Delta m^2/2E$ which is much
better suited to study flavor oscillation than energy itself.
Finally, the flux matrices are written in the form
\begin{equation}
{\bf\Phi}_{\omega,u}=\frac{{\rm Tr}\,{\bf\Phi}_{\omega,u}}{2}
+\frac{F_{\omega,u}^e-F_{\omega,u}^x}{2}\,
\begin{pmatrix}
s_{\omega,u}&S_{\omega,u}\\
S_{\omega,u}^*&-s_{\omega,u}
\end{pmatrix}\,,
\end{equation}
where $F_{\omega,u}^{e}$ and $F_{\omega,u}^{x}$ are the flavor
fluxes at the neutrino sphere. The flux summed over all
flavors, ${\rm Tr}\,{\bf\Phi}_{\omega,u}$, is conserved in our
free-streaming limit. The $\nu_e$ survival probability is
$\frac{1}{2}[1+s_{\omega,u}(r)]$ in terms of the ``swap factor''
$-1\leq s_{\omega,u}(r)\leq1$. The off-diagonal element
$S_{\omega,u}$ is complex and $s^2_{\omega,u}+|S_{\omega,u}|^2=1$.

We have formulated our equations so that, for $\Delta m^2>0$, they
describe the inverted hierarchy (IH) of the two possible neutrino
mass orderings. The normal hierarchy (NH) can be implemented with
the substitution \hbox{$\Delta m^2\to-\Delta m^2$} or, equivalently,
$\omega\to-\omega$.

\subsection{Stability condition}

The possible onset of self-induced flavor conversions is best
described in terms of the complex numbers $S_{\omega,u}$ which are
very small as long as neutrinos are in the eigenstates
of the weak interaction in the presence of matter. The small-amplitude limit means
$|S_{\omega,u}|\ll1$ and to linear order $s_{\omega,u}=1$. Assuming
in addition a large distance from the source so that $1-v_u\ll 1$,
the evolution equation linearized in $S_{\omega,u}$ and in $u$ is
\cite{Banerjee:2011fj}
\begin{eqnarray}\label{eq:smallEoM}
{\rm i}\partial_r S_{\omega,u}&=&
(\omega + u\bar\lambda)\,S_{\omega,u}
\nonumber\\
& &-\mu \int du'\,d\omega'\,(u+u')\,g_{\omega'u'}\,S_{\omega',u'}\,.
\label{stability-eom}
\end{eqnarray}
Here, $g_{\omega,u}$ is the neutrino spectrum ($\omega<0$ for
antineutrinos) which we normalize to the antineutrino flux, i.e.\
$\int_{-\infty}^0 d\omega\int_0^1 du\,g_{\omega,u}=-1$. The
``asymmetry'' between neutrinos and antineutrinos is
$\epsilon=\int d\omega\, du\,g_{\omega,u}$.

Refractive effects are encoded in the $r$-dependent parameters
\begin{eqnarray}
\lambda&=&\sqrt{2}\,G_{\rm F}\,[n_e(r)-n_{\bar e}(r)]\,\frac{R^2}{2r^2}\,,
\nonumber\\
\mu&=&\frac{\sqrt{2}\,G_{\rm F}\,[F_{\bar\nu_e}(R)-F_{\bar\nu_x}(R)]}{4\pi r^2}
\,\frac{R^2}{2r^2}\,,
\end{eqnarray}
and often combined in $\bar{\lambda} = \lambda + \epsilon\mu$. The
factor $R^2/2r^2$ means that only the multiangle impact of the
neutrino-neutrino and matter effects are relevant for our stability
analysis, not the densities themselves. We normalize
the neutrino-neutrino interaction strength $\mu$, and
consequently the spectrum $g_{\omega,u}$,
to the $\bar\nu_e$--$\bar\nu_x$ flux difference at a chosen radius
$R$, the nominal neutrino sphere. Physical results do not
depend on the choice of $R$.

Writing solutions of the linear differential equation,
Eq.~(\ref{eq:smallEoM}), in the form
\begin{equation}
S_{\omega,u}=Q_{\omega,u}\,e^{-{\rm i}\Omega r}
\end{equation}
with complex frequency $\Omega=\gamma+{\rm i}\kappa$ and eigenvector
$Q_{\omega,u}$ leads to the eigenvalue
equation~\cite{Banerjee:2011fj}
\begin{equation}\label{fourier-eom}
(\omega + u\bar\lambda - \Omega)\, Q_{\omega,u}=
\mu \int du'\,d\omega'\,(u+u')\,g_{\omega'u'}\,Q_{\omega',u'}\,.
\end{equation}
The right-hand side of this equation is a linear polynomial in $u$ so that
the eigenvector must have the form
\begin{equation}
Q_{\omega,u}=\frac{a+bu}{\omega+u\bar\lambda-\Omega}\,,
\label{eq:Eigenfunctions}
\end{equation}
where $a$ and $b$ are complex numbers. The form of the
eigenfunctions is that of a M\"obius transformation in the complex plane.
This means that for
fixed $u$ we have a circle parametrized by $\omega$ and for fixed
$\omega$ a circle parametrized by $u$. The
physical range $u\in(0,1)$ then maps to a circular arc in the
complex plane.

Following Ref.~\cite{Banerjee:2011fj} we note that, after inserting
Eq.~(\ref{eq:Eigenfunctions}) into Eq.~(\ref{fourier-eom}), both
sides are linear polynomials in $u$. Self-consistency requires
\begin{equation}\label{eq:abmatrix}
\begin{pmatrix} I_1-1&I_2\\ I_0&I_1-1\end{pmatrix}
\begin{pmatrix} a\\ b\end{pmatrix}=0\,,
\end{equation}
where
\begin{equation}
I_n=\mu \int du\,d\omega\,
\frac{u^n\,g_{\omega,u}}{\omega+u\bar\lambda-\Omega}\,.
\label{eq:In-def}
\end{equation}
In contrast with Ref.~\cite{Banerjee:2011fj} we have included a
factor $\mu$ in the integral to make $I_n$ dimensionless.

Nontrivial solutions for $a$ and $b$ exist if the determinant of the
matrix in Eq.~(\ref{eq:abmatrix}) vanishes, implying
\begin{equation}\label{eq:multiangleeigenvalue}
(I_1-1)^2=I_0 I_2\, .
\end{equation}
Changing the neutrino mass hierarchy from inverted to normal is
simply achieved by the sign change $\omega\to-\omega$ in the
denominator of the integrand of Eq.~(\ref{eq:In-def}).

\section{Continuum versus Single Angle}
\label{sec:single angle}

\subsection{Neutrino spectrum}

We use the simplest nontrivial setup that allows us to study the
role of discrete angle modes, i.e., we consider a neutrino flux
streaming from a sphere at radius $R$ that emits monochromatic
fluxes of $\bar\nu_e$ and $\nu_e$. The $\nu_e$ flux is taken to be
$1+\epsilon$ times the $\bar\nu_e$ flux, representing
deleptonization. The chosen vacuum frequency for monochromatic
neutrinos and antineutrinos $\pm\omega_0$ determines the frequency
scale of the system. We simplify the calculations by choosing
$\omega_0=1$ as the unit of measure for all other frequencies such
as $\mu$, $\lambda$, $\kappa$ and $\gamma$.

The angle distribution is taken to be black-body-like; i.e., the
neutrino sphere is taken to emit neutrinos isotropically into space
without limb darkening. This assumption corresponds to a
box spectrum in the $u$ variable of the form
\begin{equation}
B(u)=\begin{cases}1&\hbox{for $0\leq u\leq1$},\\
0&\hbox{otherwise.}\end{cases}
\end{equation}
Therefore, overall we study the neutrino spectrum
\begin{equation}\label{eq:simplespectrum}
g_{\omega,u}=\bigl[-\delta(-\omega_0-\omega)+
(1+\epsilon)\,\delta(\omega_0-\omega)\bigr]\,B(u)\,.
\end{equation}

\subsection{Inverted-hierarchy solution}

\begin{figure}[b]
\includegraphics[width=0.8\columnwidth]{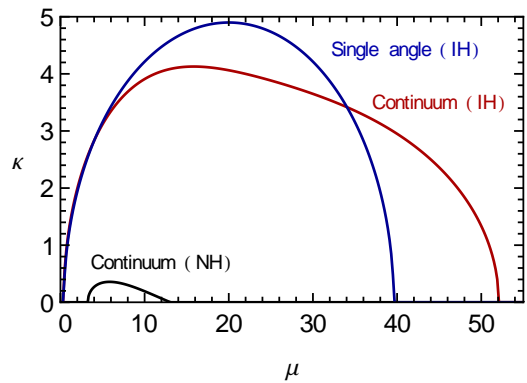}
\caption{Growth rate $\kappa$ as a function of
effective neutrino density $\mu$ in the absence of matter
($\lambda=0$) and assuming $\epsilon=1/2$. We show the single-angle
and continuum cases for normal and inverted hierarchy.
\label{fig:single1}}
\end{figure}

To solve for the eigenvalues $\Omega$ with the spectrum
Eq.~(\ref{eq:simplespectrum}) one can perform the integrals $I_n$
analytically, leading to expressions involving logarithms and the
arctan function. It is then straightforward to solve
Eq.~(\ref{eq:multiangleeigenvalue}) for the eigenvalues $\Omega$
numerically. In the absence of matter ($\lambda=0$) and for
$\epsilon=1/2$ we find the ``Continuum (IH)'' growth rate
$\kappa={\rm Im}(\Omega)$ shown in Fig.~\ref{fig:single1}. The
interacting neutrino stream is stable for $\mu$ above a critical
value $\mu_2$ and below another $\mu_1$.

It is instructive to compare this continuum result with the
single-angle approximation where we replace the box spectrum with a
single mode placed at its center: $B(u)\to\delta(1/2-u)$. This
example corresponds to the original ``flavor
pendulum''~\cite{Hannestad:2006nj}. One can solve the eigenvalue
equation explicitly and finds
\begin{equation}
\Omega=-\frac{\epsilon\mu}{2}\pm
\sqrt{1-(2+\epsilon)\mu+\left(\frac{\epsilon\mu}{2}\right)^2}\,.
\end{equation}
The solution has an imaginary part for $\mu$ between the values
$\mu_{1,2}=(1+\epsilon/2\pm\sqrt{1+\epsilon})^{-1}$. For
$\epsilon=1/2$ we find $\mu_{1,2}=20\pm4\sqrt{24}$. The
maximum growth rate and the corresponding $\mu$ value is
\begin{equation}
\kappa_{\rm max}=\frac{2\sqrt{1+\epsilon}}{\epsilon}
\quad\hbox{at}\quad \mu_{\rm max}=\frac{2(2+\epsilon)}{\epsilon^2}\,.
\end{equation}
For $\epsilon=1/2$ we find $\kappa_{\rm max}=\sqrt{24}$ and
$\mu_{\rm max}=20$. In other words, while $\omega_0$ is the only
frequency scale in our problem and the dimensionless parameter
$\epsilon$ is not especially small, the $\mu$ range where the system
is unstable reaches to $\mu_2\gg\omega_0$ and likewise, typical
$\kappa$ values are $\omega_0$ times a significant numerical factor.
A simple dimensional analysis could have suggested
$\mu\sim\kappa\sim\omega_0$.

\begin{figure}[b]
\includegraphics[width=0.8\columnwidth]{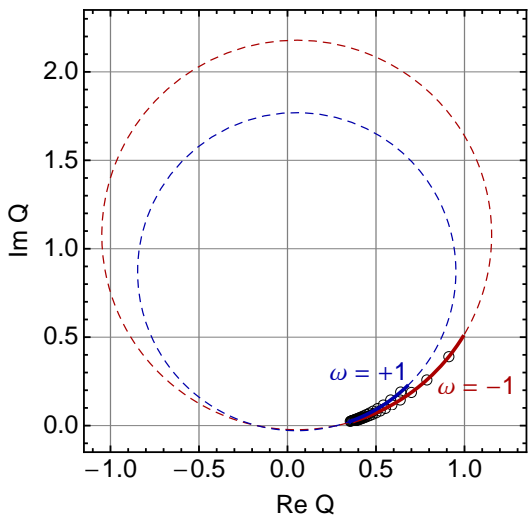}
\caption{Eigenfunction $Q_{\omega,u}$ for a box spectrum at $\mu=50$.
The solid arcs correspond to the physical range $0\leq u \leq 1$.
We also show as open circles an example for discrete angle eigenvectors
with $N_a=20$.
\label{fig:vec1}}
\end{figure}

In Fig.~\ref{fig:vec1} we show the continuum-case eigenfunctions
$Q_{\omega,u}$ for $\omega=\pm 1$, assuming $\mu=50$. As remarked
earlier, $Q_{\omega,u}$ as a function of $0\leq u\leq1$ is a
circular arc in the complex plane for fixed $\omega$. We have
normalized $Q_{\omega,u}$ so that the circle centers for different
$\omega$ lie on a vertical line in the complex plane and the
function $Q_{\omega=0,u}$ is a circle of unit radius.

The example in Fig.~\ref{fig:vec1} is close to the upper instability
range, i.e., roughly where neutrinos would start converting when the
system evolves from high to low $\mu$ values in a SN. The angular
modes remain fairly close to each other, corresponding to the
observation that the multiangle system evolves almost like the
single-angle one.

As time goes on and the unstable mode grows exponentially, the
system is described by an eigenvector such as the one shown in
Fig.~\ref{fig:vec1}, independently of the initial condition. The
evolution consists of an exponential growth with rate $\kappa={\rm
Im}(\Omega)$ away from the origin in Fig.~\ref{fig:vec1} and at the
same time a precession around the origin with frequency $\gamma={\rm
Re}(\Omega)$.

\subsection{Normal-hierarchy solution}

We next repeat this exercise for NH. Here, in the single-angle case,
the system is always stable, as already known from the flavor
pendulum~\cite{Hannestad:2006nj}. However, in the multiangle case
the system does have instabilities~\cite{EstebanPretel:2007ec}. In
Fig.~\ref{fig:single1} we show the growth rate as a function of
$\mu$ also for NH. The system is unstable only for a relatively
small $\mu$ range and the instability parameter $\kappa$ is an order
of magnitude smaller than for IH.

Thus far we have ignored the ordinary matter effect. It can suppress
self-induced flavor conversion~\cite{EstebanPretel:2008ni,
Chakraborty:2011gd, Sarikas:2011am}, and this ``multiangle matter
effect'' is particularly effective in suppressing the NH
instability. Therefore, we continue to focus on the IH case.

\section{Discrete Angle Modes}
\label{sec:fewmodes}

\subsection{Eigenvalue equation}

We next turn to our main topic, the behavior of the system
discretized in angle with $N_a>1$ angle bins, and in energy with
$N_E$ energy bins. As we always consider both neutrinos and antinautrinos, the number of frequency bins is $N_\omega = 2 N_E$. The spectrum is then implemented as
\begin{equation}
g_{\omega,u}=\sum_{i=1}^{N_\omega}\sum_{b=1}^{N_a}\,
g_{i,b}\,\delta(\omega_i-\omega)\,\delta(u_b-u)\,,
\end{equation}
leading to
\begin{equation}
I_n=\mu \sum_{i=1}^{N_\omega}\sum_{b=1}^{N_a}\,
\frac{u^n_b\,g_{i,b}\,}{\omega_i+u_b\bar\lambda-\Omega}\,.
\end{equation}
One can then determine the eigenvalues $\Omega$ by solving
Eq.~(\ref{eq:multiangleeigenvalue}) which amounts to finding the
roots of a polynomial in $\Omega$ of order $N_\omega N_a$.

Alternatively, one can begin with the eigenvalue equation of
Eq.~(\ref{fourier-eom}) in discrete form
\begin{equation}\label{fourier-eom2}
(\omega_k + u_c\bar\lambda- \Omega)\, Q_{k,c}=
\mu
\sum_{i=1}^{N_\omega}\sum_{b=1}^{N_a}\,
(u_c+u_{b})\,g_{i,b}\,Q_{i,b}\,.
\end{equation}
This equation is of the form $(M-\Omega)\,Q=0$ where $Q$ is an
$N_\omega N_a$ dimensional vector of complex numbers and $M$ a
$N_\omega N_a\times N_\omega N_a$ matrix. What remains is to find
the eigenvalues $\Omega$ and eigenvectors $Q_\Omega$ of $M$. Both
methods  provide the same eigenvalues and eigenvectors.

\subsection{Hair-comb spectrum}

\begin{figure}[b]
\includegraphics[width=0.8\columnwidth]{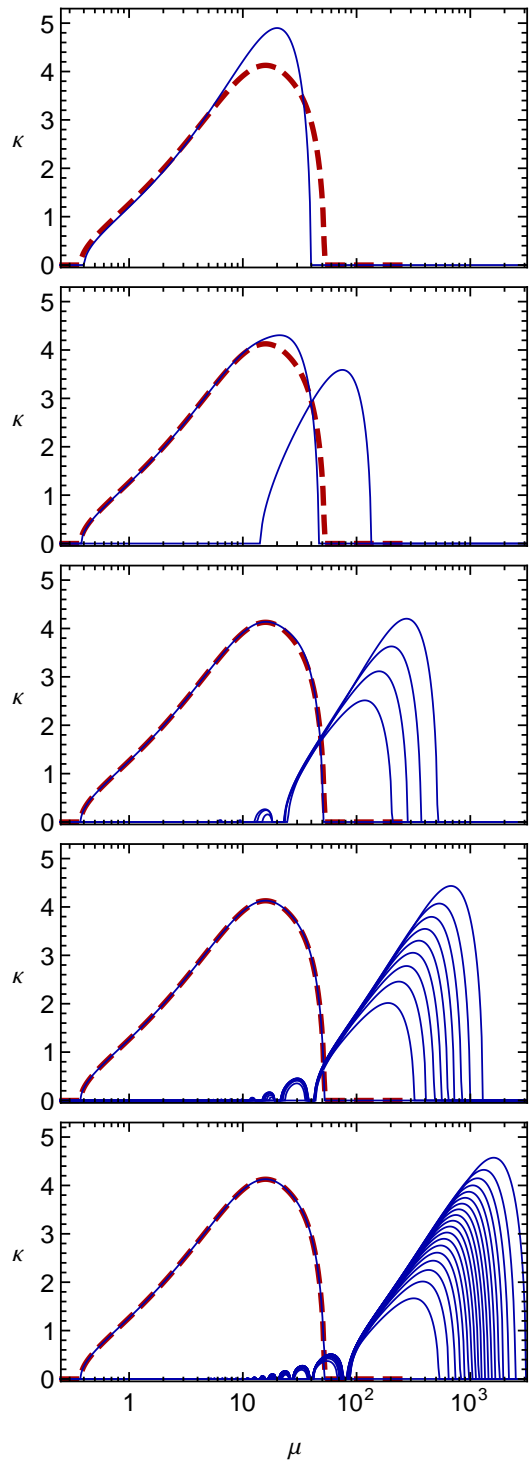}
\caption{Growth rates for unstable modes of the hair-comb
spectrum in inverted hierarchy without matter ($\lambda=0$) and $\epsilon=1/2$.
The number of angle modes is $N_a=1$, 2, 5, 10 and 20 (top to bottom).
As a red dashed line we show the continuum case.
The top panel corresponds to Fig.~\ref{fig:single1}.\label{fig:discrete1}}
\end{figure}

We will concentrate on the discrete monochromatic spectrum
($N_\omega=2$) with $N_a$ angle modes representing the original box
spectrum in the form
\begin{equation}
B(u)\to H(u)=\frac{1}{N_a}
\sum_{b=1}^{N_a}\delta\left(\frac{b-1/2}{N_a}-u\right)\,.
\end{equation}
For $N_a=1$ this is our previous single-angle case and for
$N_a\to\infty$ we expect to recover the continuum limit. We can
solve the discrete version of the eigenvalue equation in Mathematica
without problem and we show the spectrum of growth rates in
Fig.~\ref{fig:discrete1}.

We always find the ``ordinary mode'' which, for large $N_a$,
approaches the continuum solution (red dashed line in
Fig.~\ref{fig:discrete1}). In addition, we find $N_a-1$
``extraordinary modes'', which arise at larger $\mu$. With
increasing $N_a$, the extraordinary modes shift their instability
regions to larger $\mu$ values and, in the limit $N_a\to\infty$,
disappear at infinity. Of course, for any finite $N_a$, the
extraordinary modes exist at a sufficiently large $\mu$.

In numerical studies, the neutrino flavor content is evolved along
the radial SN direction, from large to small $\mu$ values. A
numerical integration must begin at a depth where the ordinary mode
is stable, i.e., to the right of the red dashed curve in
Fig.~\ref{fig:discrete1}. Inevitably, the system first encounters
the extraordinary instabilities, leading to spurious solutions.
Therefore, for a chosen inner boundary radius $r_0$, the number of
angle modes $N_a$ must be large enough so that the extraordinary
instabilities disappear to depths below it. If $r_0$ is chosen
closer to the neutrinosphere, implying a larger $\mu$, the required
$N_a$ is even larger. The most economical choice for the inner
boundary radius $r_0$ is at the large-$\mu$ beginning of the
ordinary instability region.

\subsection{Nature of the extraordinary modes}

The presence of more than one unstable mode is not surprising:
solving Eq.~(\ref{eq:multiangleeigenvalue}) for a discrete case
leads to a polynomial in $\Omega$ of order $2 N_a$ and to equally
many eigenvalues, some or all of which can have an imaginary part.
However, empirically the extraordinary modes are quite distinct
from the ordinary one.

\begin{figure}[b]
\includegraphics[width=0.8\columnwidth]{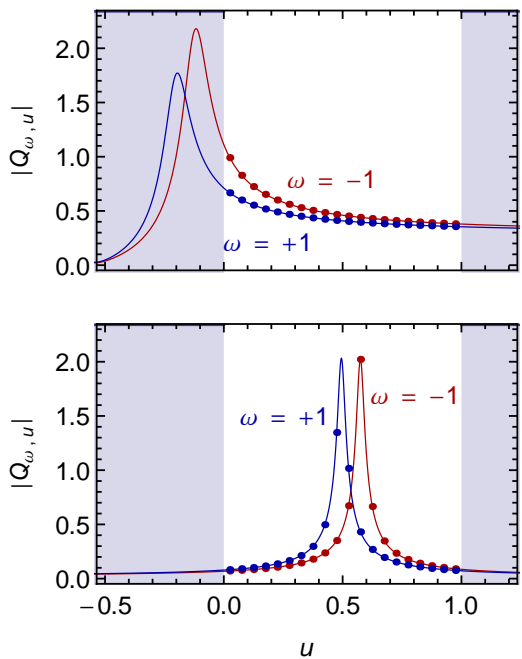}
\caption{Modulus of the eigenvector $Q_{\omega,u}$, according to
Eq.~(\ref{eq:Eigenfunctions}),
for $\mu=50$ and the hair-comb angle distribution with $N_a=20$.
Only the range $0\leq u\leq 1$ is physical.
Top: Ordinary mode. Bottom: The extraordinary mode with the largest
$\kappa$.\label{fig:disvec}}
\end{figure}

One difference can be gleaned from the general form of the
eigenvector $Q_{\omega,u}$, Eq.~(\ref{eq:Eigenfunctions}), and especially the
resonance denominator $\omega+u\bar\lambda-\gamma-\I\kappa$. Varying
$u$ between 0 and 1 over a continuous or discrete set of values may
lead one to encounter a resonance of approximate width
$\kappa/\bar\lambda$. We illustrate this point in
Fig.~\ref{fig:disvec} where we show an example for the modulus of
$Q_{\omega,u}$ as a function of $u$; the unshaded range corresponds
to the physical range $0\leq u\leq 1$. For the ordinary mode, the
resonance lies in the unphysical range, implying that
$|Q_{\omega,u}|$ does not vary much as a function of $u$. In other
words, all angle modes are close to each other and evolve similar to
the single-angle case. Avoiding the resonance is possible if the
precession frequency is negative $\gamma<0$, and this is indeed the
case for the ordinary (or quasi-single-angle) mode.

The extraordinary modes, on the other hand, always seem to have
$\gamma>0$, i.e., they all precess in the opposite direction. The
resonance falls in the physical $u$ range and one or a few angle
modes have $|Q_{\omega,u}|$  much larger than the others. In a plot
of the circular arcs as in Fig.~\ref{fig:vec1}, the ordinary mode
for $0\leq u \leq1$ traces out a small part of the circle, but the
extraordinary ones trace most of the circle. For our hair-comb spectra,
the width of the resonance as a function of $u$ has the approximate
width of the spacing of the discrete $u$ modes; i.e., typically one
or two angle modes are on resonance. The different extraordinary
modes differ in the angle mode that is on resonance. By their very
nature, these modes are not quasi-single-angle and it is
unsurprising that, when they have grown beyond the linearized
regime, they tend to cause kinematical decoherence among angle
modes.

We stress that, as far as the stability analysis is concerned, going
beyond one energy bin (two frequencies) is not necessary. Since the
equations of motion, Eq.~(\ref{eq:smallEoM}), are linear in
frequency $ \omega$, an average frequency can represent the whole
spectrum. A multienergy treatment does provide, of course, more
eigenvalues with respect to the monochromatic study, but no
additional complex ones. The number of possible unstable solutions
is related to the number of ``spectral crossings''
\cite{Dasgupta:2009mg}. The additional solutions introduced by
additional energy bins are purely real, and as such have no
importance in the stability analysis. A multienergy treatment is
necessary only in simulations if one is to resolve spectral
features.

\subsection{Other cases of extraordinary modes}

Extraordinary modes are not limited to discrete angle
distributions. For example, a $u$ spectrum consisting of two boxes
(instead of two delta spikes) has one ordinary and one
extraordinary mode. Two boxes are equivalent to a single box with a
gap, and the $\mu$ range where the extraordinary mode appears
migrates to larger $\mu$ values as the gap is chosen to be narrower.
Likewise, if we consider two delta spikes with a separation $\Delta
u$, then the solution for $\Delta u=1/2$ corresponds to the upper
panel of Fig.~\ref{fig:discrete1}, $\Delta u=0$ corresponds to the single-angle
case, and for any other value one obtains two solutions, the
extraordinary one migrating to higher $\mu$ values for decreasing
$\Delta u$.

Generally it appears that ``sharp features,'' notably steps, in the
angle spectrum cause extraordinary modes. (Of course we mean the following:
extraordinary modes with nonvanishing growth rate $\kappa>0$.)
However, it happens only for the ascending steps (for increasing $u$),
and not for the descending ones. For example, a descending staircase
spectrum has only the
ordinary (quasi-single-angle) mode, while an ascending one has as many
extraordinary modes as steps minus 1. If the steps are somewhat
smoothed, we still get the extraordinary modes. The location of the
extraordinary mode on the $\mu$ axis depends on how narrow the
spectral feature is, and the maximum $\kappa$ depends on the magnitude of
the jump.

A more mathematical classification of these observations is not
available at present. In a realistic SN situation, the continuous
angle spectrum is not a box, but typically a smoothly varying broad
distribution. As such it should not have any extraordinary modes,
or at least none with $\kappa$ values comparable to the ordinary
mode. In realistic numerical SN simulations, the appearance of
extraordinary modes is probably caused only by a discretized angle
spectrum.

\section{Multiangle matter suppression}

\begin{figure}[ht]
\includegraphics[width=0.8\columnwidth]{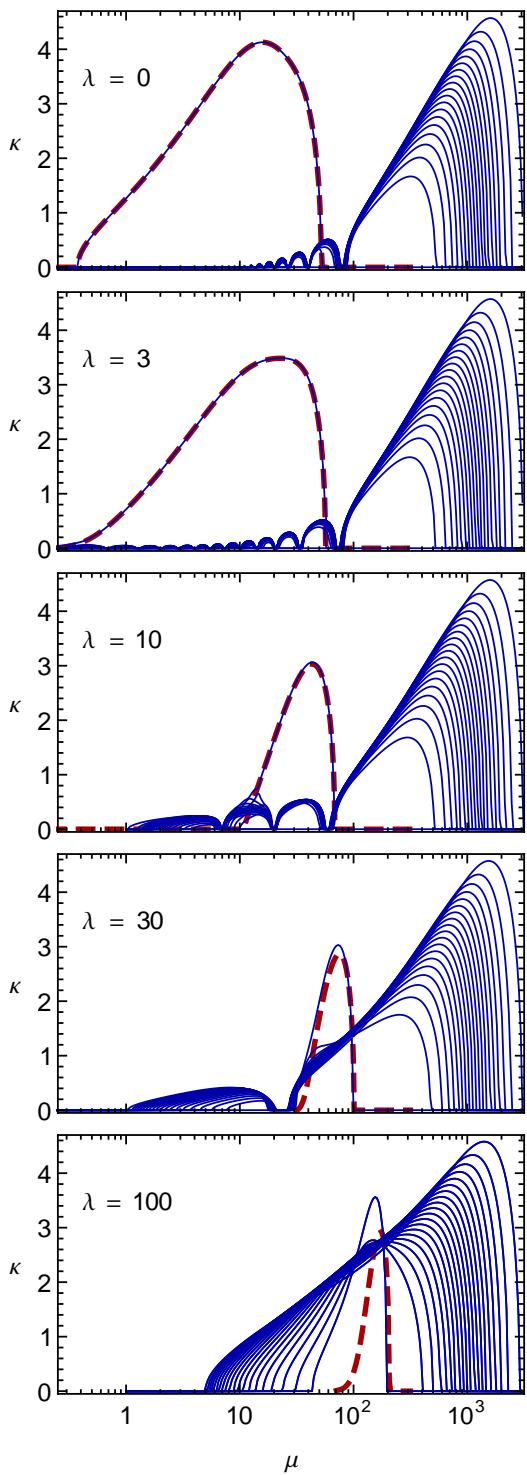}
\caption{Growth rates using the indicated values $\lambda$ for
the matter effect. Red dashed linea: Continuous box spectrum for angles.
Blue solid lines: Hair-comb representation with $N_a=20$.
\label{fig:discrete2}}
\end{figure}

It is well known that a large density of matter, represented by the
parameter $\lambda$, strongly modifies the instability in a
multiangle treatment \cite{EstebanPretel:2008ni, Raffelt:2008hr,
Duan:2010bf, Chakraborty:2011gd, Sarikas:2011am}. This ``multiangle
matter effect'' does not appear in the single-angle case and is one
important motivation to go beyond single-angle studies in the first
place. Typically, nonvanishing growth rates appear only for
$\mu\sim\lambda$ \cite{Banerjee:2011fj, Sarikas:2011am}. In other
words, for $\mu\alt\lambda$ the instability is suppressed and
collective flavor conversion may not occur, notably during the
accretion phase~\cite{Chakraborty:2011gd, Sarikas:2011am}.

To understand the impact of matter in the context of a discrete
angle spectrum, we consider once more a box spectrum of angles. In
Fig.~\ref{fig:discrete2} we show, as a red dashed line, the growth
rate for different values of $\lambda$. We see that indeed the
low-$\mu$ instability is suppressed and that the instability region
is shifted to $\mu\sim\lambda$.

In addition, we show the solutions for a hair-comb representation
with $N_a=20$. For the extraordinary modes, the presence of
$\lambda$ has somewhat the opposite effect of enhancing the growth
rates at low $\mu$ values. In the bottom panel of
Fig.~\ref{fig:discrete2} we see that the ordinary mode disappears in
the forest of extraordinary ones. In other words, in the presence
of matter one needs a yet larger value for $N_a$ to shift these
modes away.

\section{Discussion}

We have used the method of linearized stability analysis to shed
light on the appearance of spurious solutions in numerical studies
of nonlinear neutrino flavor evolution. Spurious solutions appear
when the neutrino radiation field is represented by discrete modes
and the number of angles $N_a$ is too small. The physical solution
tends to be one where the angle modes remain nearly aligned (quasi
single angle) and which we have called the ordinary mode. In
addition, extraordinary modes appear at values of the effective
neutrino density $\mu$ that depends on the spacing of the discrete
modes. If they are sufficiently densely spaced, the extraordinary
modes are at large $\mu$ (small SN radius) such that they do not
spoil a numerical multiangle simulation.

The linear stability approach allows one to determine, without much
effort, the growth rate of the physical instability as a function of
the SN radius. Any instability is important only if the growth rate is
large enough to take the system into the nonlinear regime on the
available length scale. In a SN core, the effective neutrino and
matter densities vary as power laws so that the relevant length
scales correspond approximately to $r$. Typical growth rates
$\kappa$ are a few times the vacuum oscillation frequency
$\omega_0$. For SN neutrino oscillations driven by the atmospheric
mass difference, typically we have $\omega_0\sim 0.5$~km$^{-1}$ and so
typical $\kappa$ values are few inverse kilometers. The available length
scales are tens to hundreds of kilometers, so the physical instability has
enough time to become nonlinear, in agreement with numerical
studies. Of course, the number of $e$-foldings required for a mode
to become nonlinear depends on its initial amplitude.

In principle, then, for a concrete numerical example it is enough to
find $\kappa(r)$ for the physical mode, based on a continuous angle
distribution, and in this way find the onset radius of the
instability. It would be enough to start the numerical integration
at that radius. Since the starting point of flavor conversions is an
exponential runaway, nothing is
gained by starting deeper, i.e., in the stable regime of the
ordinary mode. The number of angle modes then has to be chosen large
enough that the growth rate of the extraordinary modes is much
smaller than that of the ordinary one in in the onset region.
Starting the integration at a smaller radius requires enough angle
modes to avoid the extraordinary modes becoming nonlinear before
the system has reached the physical onset radius.

If one determines the required $N_a$ by trial and error for a given
numerical example, it has been observed that the solution becomes
reproducible for $N_a$ above some critical value, and nothing much
changes by choosing $N_a$ yet larger. This behavior corresponds to
the aforementioned requirement that the extraordinary modes must not
become nonlinear before the physical onset radius. The solution then
no longer changes because, at the onset radius, the system will
always select the physical eigenvector from whichever configuration
the neutrino ensemble is in, by subjecting it to the exponential
growth that will allow it to dominate the final outcome.

The appearance of unphysical modes in the discretized problem
suggests that one is using a bad representation of the physical
system. It would be desirable to cast the original equations of
motion in a form that allows for a numerical treatment while
avoiding spurious solutions.

Instead of using discrete angles one may consider spherical
harmonics~\cite{Raffelt:2007yz}. We have attempted this approach,
truncating the expansion at some multipole order. Nevertheless, we
obtain similar extraordinary modes and the required number of
polynomials corresponds approximately to the required number of
discrete angles. It remains to be seen whether a suitable closure of the
multipole equations of motion can be found that avoids unphysical
solutions without going to very high multipole order. Conversely, if
this is not possible, it would be important to understand whether
extraordinary modes are unavoidable in any scheme that fails to
resolve sufficiently fine details of the angle distribution.

For now the question remains open if numerical brute force is the
only way forward to understand SN neutrino flavor evolution in more
general situations beyond the toy models that have been studied thus
far, or if the equations can be set up in a way that avoids the need
for excessive computer power.

\section*{Acknowledgements} 

This work was partly supported by the Deutsche
Forschungsgemeinschaft under Grant No. EXC-153 (Cluster of Excellence
``Origin and Structure of the universe'') and by the European Union
under Grant No. PITN-GA-2011-289442 (FP7 Initial Training Network
``Invisibles''). D.S. acknowledges support by the
Funda\c{c}\~{a}o para a Ci\^{e}ncia e Tecnologia (Portugal).



\begin{thebibliography}{00}

\bibitem{Duan:2006an}
  H.~Duan, G.~M.~Fuller, J.~Carlson and Y.-Z.~Qian,
  Phys.\ Rev.\  D {\bf 74}, 105014 (2006).

\bibitem{Duan:2010bg}
  H.~Duan, G.~M.~Fuller and Y.-Z.~Qian,
  Annu.\ Rev.\ Nucl.\ Part.\ Sci.\ {\bf 60}, 569 (2010).

\bibitem{Duan:2008eb}
  H.~Duan, G.~M.~Fuller and J.~Carlson,
  Comput.\ Sci.\ Dis.\  {\bf 1}, 015007 (2008).

\bibitem{Raffelt:2007cb}
  G.~Raffelt and A.~Yu.~Smirnov,
  Phys.\ Rev.\  D {\bf 76}, 081301 (2007);
  {\bf 77}, 029903(E) (2008);
%
  Phys.\ Rev.\ D {\bf 76}, 125008 (2007).

\bibitem{Duan:2007fw}
  H.~Duan, G.~M.~Fuller and Y.-Z.~Qian,
  Phys.\ Rev.\  D {\bf 76}, 085013 (2007).

\bibitem{Fogli:2007bk}
  G.~L.~Fogli, E.~Lisi, A.~Marrone and A.~Mirizzi,
  J. Cosmol. Astropart. Phys 12 ({\bf 2007}) 010.
%
  G.~L.~Fogli, E.~Lisi, A.~Marrone, A.~Mirizzi and I.\ Tamborra,
  Phys.\ Rev.\  D {\bf 78}, 097301 (2008).

\bibitem{Dasgupta:2009mg}
  B.~Dasgupta, A.~Dighe, G.~Raffelt and A.\ Yu.\ Smir\-nov,
  Phys.\ Rev.\ Lett.\  {\bf 103}, 051105 (2009).

\bibitem{Hannestad:2006nj}
  S.~Hannestad, G.~Raffelt, G.~Sigl and Y.~Y.~Y.~Wong,
  Phys.\ Rev.\ D {\bf 74}, 105010 (2006);
  {\bf 76}, 029901(E) (2007).

\bibitem{Duan:2007mv}
  H.~Duan, G.~M.~Fuller, J.~Carlson and Y.-Z.~Qian,
  Phys.\ Rev.\ D {\bf 75}, 125005 (2007).

\bibitem{EstebanPretel:2007ec}
  A.~Esteban-Pretel, S.~Pastor, R.~Tom\`as, G.~G.~Raffelt and G.~Sigl,
  Phys.\ Rev.\ D {\bf 76}, 125018 (2007).

\bibitem{Raffelt:2010za}
  G.~G.~Raffelt and I.~Tamborra,
  Phys.\ Rev.\ D {\bf 82}, 125004 (2010).

\bibitem{Raffelt:2011yb}
  G.~G.~Raffelt,
  Phys.\ Rev.\ D {\bf 83}, 105022 (2011).

\bibitem{Yuzbashyan:2008}
  E.~A.~Yuzbashyan,
  Phys.\ Rev.\ B {\bf 78}, 184507 (2008).

\bibitem{Pehlivan:2011hp}
  Y.~Pehlivan, A.~B.~Balantekin, T.~Kajino and T.~Yoshida,
  Phys.\ Rev.\ D {\bf 84}, 065008 (2011).

\bibitem{Sawyer:2008zs}
  R.~F.~Sawyer,
  Phys.\ Rev.\  D {\bf 79}, 105003 (2009).

\bibitem{Banerjee:2011fj}
  A.~Banerjee, A.~Dighe and G.~Raffelt,
  Phys.\ Rev.\ D {\bf 84}, 053013 (2011).

\bibitem{Sarikas:2011am}
  S.~Sarikas, G.~G.~Raffelt, L.~H\"udepohl and H.-T.~Janka,
  Phys.\ Rev.\ Lett.\  {\bf 108}, 061101 (2012).
  S.~Sarikas and G.~Raffelt,
  arXiv:1110.5572.
  S.~Sarikas, I.~Tamborra, G.~Raffelt, L.~H\"udepohl and H.-T.~Janka,
  Phys.\ Rev.\ D {\bf 85}, 113007 (2012).

\bibitem{Chakraborty:2011gd}
  S.~Chakraborty, T.~Fischer, A.~Mirizzi, N.~Saviano and R.~Tom\`as,
  Phys.\ Rev.\  D {\bf 84}, 025002 (2011);
  Phys.\ Rev.\ Lett.\  {\bf 107}, 151101 (2011).
  N.~Saviano, S.~Chakraborty, T.~Fischer and A.~Mirizzi,
  Phys.\ Rev.\ D {\bf 85}, 113002 (2012).

\bibitem{EstebanPretel:2008ni}
  A.~Esteban-Pretel, A.~Mirizzi, S.~Pastor, R.~Tom\`as,
  G.~G.~Raffelt, P.~D.~Serpico and G.~Sigl,
  Phys.\ Rev.\  D {\bf 78}, 085012 (2008).

\bibitem{Raffelt:2008hr}
  G.~G.~Raffelt,
  Phys.\ Rev.\ D {\bf 78}, 125015 (2008).

\bibitem{Duan:2010bf}
  H.~Duan and A.~Friedland,
  Phys.\ Rev.\ Lett.\  {\bf 106}, 091101 (2011).

\bibitem{Raffelt:2007yz}
  G.~G.~Raffelt and G.~Sigl,
  Phys.\ Rev.\  D {\bf 75}, 083002 (2007).

\end{thebibliography}
\end{document}